\documentclass[
superscriptaddress,
 nofootinbib,
 amsmath,amssymb,
11pt,
]{revtex4-2}

\usepackage[utf8]{inputenc}
\usepackage{subcaption}
\usepackage{graphicx}
\usepackage{dcolumn}
\usepackage{bm}
\usepackage{needspace}
\usepackage{xcolor}
\usepackage{wrapfig}
\usepackage[font=scriptsize,labelfont=bf]{caption}
\usepackage{floatrow}
\usepackage{verbatim}

\usepackage[colorlinks=True, citecolor=blue, urlcolor=blue]{hyperref}
\usepackage{booktabs}
\usepackage{natbib}

\begin{document}
\title{Quantum computers as an amplifier for existential risk}%
\author{Benjamin~F.~Schiffer}
\email[Correspondence: ]{Benjamin.Schiffer@mpq.mpg.de}
\affiliation{\small Faculty of Philosophy, Philosophy of Science and the Study of Religion, Ludwig-Maximilians-Universität, D-80539 Munich, Germany }%
\date{\today}

\begin{abstract}
Quantum computing is expected to have a profound impact on society. In this work we discuss the potential consequences on existential risk for humanity. Even with the timeline for large-scale fault-tolerant quantum computing still unclear, it is highly likely that quantum computers will eventually realize an exponential speedup for certain practical applications. We identify quantum simulation as the most relevant application in this regard and we qualitatively outline different risk trajectories. Both amplifying and mitigating effects of quantum computing for existential risk are anticipated. In order to prevent quantum computing from being an amplifier of existential risk, we call for increased efforts by the scientific community towards reducing potential future quantum risk. This viewpoint seeks to add a new perspective to the discussion on technological risk of quantum computing.
\end{abstract}
\maketitle

\section{Introduction}
In recent years, we witnessed remarkable progress in the field of quantum computing both in theoretical foundations as well as in experimental realizations. There have been first instances of computational problems where quantum devices were able to compute the solution faster than any classical supercomputer - thereby achieving so-called quantum advantage \cite[]{Arute2019Quantuma, Zhong2020Quantuma}. 
While the first efforts to imagine useful quantum computing started in research groups around the world about 40 years ago, this collective effort is now shared by strong industry players such as Google, IBM, AWS, Microsoft or Alibaba, who mostly joined the field in the last decade.\\
While public perception of quantum computing seems to be embracing this early technology quite enthusiastically, scientist repeatedly need to adjust the high expectations placed on quantum computing. Here, we are going to take a somewhat different perspective. We would like to pose the question what would be the consequences for humanity if we indeed succeed in building a large scale fault-tolerant quantum computer. Our question is motivated by a similar discussion about emerging risk from artificial general intelligence. So far, only few scholars have been addressing the technology implications of quantum computing \cite[]{Wolf2017potential} and even less so in the context of existential risk. This viewpoint is an attempt to close this gap and builds upon the concept of existential risk as presented by Ord~\cite[]{Ord2020}.
\\
In this work, we shall argue that there is an existential threat to humanity arising from the prospect of being able to run quantum simulation on a quantum computers in the future. 
The structure of our argument is as follows: First, we discuss what quantum computers are and argue that large-scale quantum computers are likely going to be realized in the next decades (Sec.~\ref{sec:qc}). 
In the next section (Sec.~\ref{sec:app}), we introduce the concept of a quantum speedup, outline key algorithms which provide an exponential speedup and identify digital quantum simulation (DQS) as the most relevant application for our discussion. Following this, we discuss the impact of DQS and argue why it has the potential to increase existential risk (Sec.~\ref{sec:impact}). Finally, we briefly discuss our findings and call for risk mitigation research in quantum computing.~(Sec.~\ref{sec:conclusion}).
\section{The road to fault-tolerant quantum computing} \label{sec:qc}
\subsection{A definition of quantum computing}
A \emph{quantum computer} harnesses the laws of quantum mechanics to perform computations. This is to be understood that for a quantum computer, quantum mechanics is not only required for the fabrication process as it is the case with modern classical computers. Instead, quantum computers execute quantum algorithms that rely intrinsically on quantum mechanical phenomena such as superposition and entanglement. The promise of quantum computing lies in the asymptotic speedup over classical algorithms for certain computational problems. This asymptotic behavior is crucial for our analysis and we will discuss it in more detail below. \\
The basic unit of quantum information is called a quantum bit (qubit) and quantum computers comprise a large number of coherently interconnected qubits. Due to the no-cloning theorem, quantum states cannot be copied, in general. This poses to be a major challenge when dealing with noise that arises occasionally in the computation. Quantum computers can be made fault-tolerant through quantum error correction which requires a large overhead in the number of qubits~\cite[]{Terhal2015Quantum}.
For the scope of this paper, we will use the term quantum computer to refer to large-scale fault-tolerant quantum computers of more than 1,000 logical (error-corrected) qubits.
\subsection{The feasibility of building large quantum computers}
We discuss the likelihood that a (large-scale) quantum computer is successfully going to be built in the future. There are two aspects that we shall consider. First, there needs to be the actual physical possibility of constructing a large-scale fault-tolerant quantum computer. Second, realizing this technology requires a sufficient economical or political interest.\\
Concerning the first point, there are, in fact, currently no physical limitations known regarding the scale-up of quantum computers. To this day, however, the problem of scaling up has not been solved due to massive engineering challenges which need to be overcome first~\cite[]{krinner2019engineering}. The fact, that these unsolved engineering challenges exist, does not change the fact that no physical limitations are expected and a large-scale quantum computer seems indeed within the realm of the possible.\\
Not every technology which is physically possible is actually being realized due to the required financial investment into the research and the realization phase. Regarding quantum computers, we observe that there is not just a political desire for its realization. Importantly, there is currently a huge industrial interest developing which comes from the prospect of a profitable commercialization of this technology.
Quantum computers are expected to speed up certain computational tasks such as optimization problems and aid in the development of new materials and chemical reactions.  Besides the industrial interest, there is also a scientific motivation to build a quantum computer. It is not only a remarkable fundamental physics experiment on its own, but might also allow new discoveries in fundamental science, e.g.~regarding exotic phases of matter or quantum gravity. The European Organization for Nuclear Research (CERN) serves as a prime example that, given sufficient interest for fundamental science, even very costly large scale scale experiments have been realized even without immediate commercial relevance. These are several reasons why the eventual realization of a large quantum computers seems highly likely.
\subsection{A timeline for quantum computing}
This leaves us with the question when we might expect such large-scale quantum computers. If one assumes that the number of qubits will increase exponentially in the next decades, 1,000 logical qubits can be expected with 50\% probability in the years 2040s by interpolating the historical development into the future~\cite[]{Sevilla2020Forecasting}. Given that there is very few data points to extrapolate from and that significant engineering challenges need to be overcome, this interpolation is to be interpreted cautiously.
Another source that needs to be treated with the appropriate prudence are the publicly announced roadmaps from quantum computing companies. Google and IonQ roughly align in their aim for 1,000 logical qubits around 2030 \cite[]{ShanklandQuantum, Scaling}. Also, the Quantum Manifesto which paved the way for the Quantum Flagship of the European Union aims for universal quantum computing around 2035~\cite[]{member_states_2016}. However, we would like to stress again the fact that key engineering challenges in the scale-up remain unsolved, adding significant uncertainty to the timeline on large-scale fault-tolerant quantum computing.\\
For our argument, the uncertainty in the timeline is not crucial. While it seems rather likely that a quantum computer will be built in this century, our arguments still apply if due to some setback quantum computers will only arrive several decades later.\footnote{In the history of artificial intelligence (AI), there was the period of an AI winter in the 1990ies when research funding and general interest in AI was at a low after expectations had failed to materialize. It is possible that some of the current expectations placed on quantum computing similarly exceed the pace of scientific progress, potentially triggering a Quantum winter.} On the contrary, one might even argue that the first experiments which showcased quantum advantage happened earlier than many in the research community would have expected. We might draw an analogy to the AlphaGo moment in machine learning, where a generally unexpected breakthrough changed the general perception of the capabilities of this technology~\cite[]{silver2016mastering}. Indeed, the first experiments which showcased quantum advantage happened earlier than many scholars in the research community would have expected. \\ 
The bottom line is that a consideration of the technological risk of quantum computing is relevant even if the commercial timelines put forward so far turn out to be on the optimistic side. However, if the arrival of large-scale quantum computing comes early due to an unexpected, sudden breakthrough, that is within the next 10-15 years already, then, risk analysis and possible mitigation would become even more urgent.
\section{Application of quantum computers} \label{sec:app}
\subsection{Quantum speedups}
Using a quantum computer to solve mathematical problems is promising if the problem is solved more time- or memory-efficient than with a classical computer. A speed difference between algorithms is mainly relevant for large problem instances and quantum speedup is defined as the ratio between the time required on a classical computer and a quantum computer as a function of the problem size~\cite[]{Roennow2014Defining}. The largest separation between quantum and classical computing is found if the quantum speedup has an exponential scaling in the problem size. For several other applications a polynomial quantum speedup has been established that might become relevant with the arrival of fault-tolerant quantum computing~\cite[]{babbush2021focus}.\\
We argue here that the disruptive impact of quantum computing is greatest where the quantum computer can reduce the running time of the algorithm so much that a previously intractable problem can then be solved in reasonable time.\footnote{What a \emph{reasonable} time is will be highly dependent on the computational problem and might as well range from few seconds to several weeks. Hence, the term is left rather vague here.} This corresponds to an exponential quantum speedup where an exponentially long classical runtime can be reduced to a polynomial quantum runtime. In this case, quantum computing has the role of an enabler, such that it becomes a tool or method which makes it possible to computationally solve a certain problem which was before not solvable in practice. Labeling certain use-cases of quantum computing as an  enabler is helpful in assessing the impact of this new technology. While gradual improvements of technology may certainly have a large impact as well, the impact can come much more sudden with an enabler-type technology. This is why we focus on applications of quantum computers where they show precisely this behavior of an enabler.
\subsection{Quantum algorithms with an exponential speedup}
In fact, there exist few quantum algorithms which provide such an enabler-type, (sub)exponential quantum speedup\footnote{A sub-exponential speedup is smaller than exponential, yet larger than polynomial. For our argument, we will not distinguish between an exponential speedup and a sub-exponential speedup.}, such as Shor's algorithm \cite[]{Shor1999Polynomial}, quantum walks on a tree graph~\cite[]{Childs2003Exponential}, the HHL algorithm~\cite[]{Harrow2009Quantum}, and digital quantum simulation (DQS)~\cite[]{Georgescu2014Quantum}. It would be beyond the scope of this essay to review all of these algorithms. Instead, we would like to comment briefly on Shor's algorithm due to its prominence, and more in detail on DQS because of the implications of this algorithmic method for our argument.
\subsubsection{Shor's algorithm}
Shor's algorithm solves the computational problem of factorization on a quantum computer. When given some integer number, finding its prime factors is a task that requires a time which scales sub-exponentially in the number of digits on a classical computer. At the same time, verifying that a set of given prime factors divides an integer, is very efficient. This asymmetry between multiplying and factoring is at the heart of the RSA cryptography method~\cite[]{Rivest1978Method}. Many services of the modern Internet rely on RSA for their encryption such as VPN, online banking, credit cards, e-mail and online shopping.\footnote{In fact, elliptic-curve cryptography (ECC) is also used widely today. However, ECC has the same weakness towards quantum computing attacks as RSA. Thus, the  discussion on breaking RSA with Shor's algorithm also applies to ECC.} Remarkably, Shor's algorithm provides an efficient method to solve prime factorization on a quantum computer and is an example of a (sub)exponential quantum speedup. While this is becoming highly problematic for the mentioned cryptographic methods, there exist other forms of encryption, e.g.~lattice-based cryptography~\cite[]{Micciancio2009Lattice}, which are assumed to be safe against attacks from quantum computers. While changing the encryption method might lead to additional implementation costs, it currently seems unlikely that Shor's algorithm has a significant impact onto existential risk in this regard.
\subsubsection{Digital quantum simulation (DQS)}
Digital quantum simulation (DQS) also provides an exponential quantum speedup and could have more serious consequences. We shall explain this application of a quantum computer and implications thereof in the following paragraphs.\\
The theory of quantum mechanics has been tremendously successful in describing nature. Since over 100 years it is being investigated, understood and challenged to ever greater extend and it has been proven to be highly accurate. Unfortunately, for large quantum mechanical systems such as molecules this fundamental theory alone is not terribly useful. This is because the number of parameters, which are needed to be kept track of, scales exponentially with the size of the quantum system. For this reason, classical computers are generally not able to (efficiently) describe large quantum systems.\\
A quantum device, however, offers the promising advantage to simulate other quantum systems efficiently. Or, as Richard Feynman famously put it, ``nature isn't classical, dammit, and if you want to make a simulation of nature, you'd
better make it quantum mechanical''~\cite[]{Feynman2002Simulating}. Quantum simulation is the study of quantum systems to understand certain properties of the system, that we might be interested in. It offers an exponential quantum speedup in understanding reaction mechanism in molecules and probing the properties of new materials. This is of great interest for the development in many branches of industry due to the very high research and development cost that might be saved.\\ Here, we shall distinguish two different approaches to quantum simulation, analogue quantum simulation and digital quantum simulation. Analogue quantum simulation aims at understanding a quantum-mechanical problem by building another quantum system with similar properties in a laboratory. This technique is being successfully employed in many different research groups and has led to the discovery and a better understanding  of new physical phenomena~\cite[]{schreiber2015observation, turner2018weak}. Generally speaking, in an analogue quantum simulation experiment, the Hamiltonian describing the physical system is fixed to a certain degree. Hence, analogue quantum simulation does not offer full flexibility in changing the quantum system of interest. For a detailed discussion on the capabilities of analogue quantum simulation, we refer to the roadmap laid out in~\cite[]{altman2021quantum}.\\
Digital quantum simulation (DQS) offers a more flexible way to perform quantum simulation being run an a programmable quantum computer. We note, however, that it inflicts a computational overhead to due the digital nature of the quantum simulation. Hence, this algorithm will only be able to solve useful problems once very large quantum computers become available with a high logical qubit count. As an example, in order to analyze a key molecule in biological nitrogen fixation, hundreds of logical qubits are required~\cite[]{Reiher2017Elucidating}. Once such quantum computers have arrived, it can be expected that research and development pipelines in industry can be dramatically shortened. Using DQS, many chemical reactions or materials could be tested virtually on a quantum computer instead of costly and time-intensive experiments. In this sense, DQS will be an enable for boosting technological development. 
\section{Impact of digital quantum simulation on existential risk} \label{sec:impact}
In the following paragraphs, we shall analyze the implications of digital quantum simulation (DQS) on technological risk, more precisely existential risk. We will give three arguments why DQS could act as an amplifier for existential risk: one general argument and two more concrete scenarios. Also, we point out how DQS might be useful to mitigate existential risk, which serves as a counter-argument.
\subsection{DQS as a general booster for technological progress}
The first argument is a general argument about the existential risk of technology. This argument builds upon the notion that the level of existential risk in the 21st century is unsustainable in the sense that humanity would with significant probability go extinct if this risk level is not reduced~\cite[]{Ord2020}. In order to reach a sustainable level of existential risk which allows for the long-term survival of humanity, the level of existential risk needs to be reduced. There are both natural causes, such as supervolcanoes or asteroids, and anthropogenic causes of existential risk. The latter include nuclear war, climate change, engineered pandemics and misaligned AI. 
While assigning probabilities to these individual risks is difficult, it is instructive to at least estimate these numbers. In fact, the likelihood for an event that brings human extinction or at least general societal collapse within the next 100 years through a human-made catastrophe is estimated multiple orders of magnitude higher than through a natural catastrophic event by Ord~\cite[p.~167]{Ord2020}. Anthropogenic existential risk was introduced through the means of modern technology. Therefore, the largest lever to lower anthropogenic existential risk is the ability for humanity to effectively and responsibly control technology. However, this poses to be a immense collective challenge and it should be expected that this process could take up to decades (cf. \cite[pp.~187-216]{Ord2020}).\\
Large-scale Quantum computing is a new technology which has a realistic chance to become first available towards the middle of the 21st century, as we argued in the previous section. It is unclear whether humanity will have gained a better understanding of how to control potentially risky technology at that point. DQS has the potential to accelerate technological development because of its ability to speed up research and development in those areas that would benefit from performing analyses of large quantum mechanical systems. This includes a broad range of applications for chemical processes, new materials etc. A broad-scale acceleration of technology development without an improvement of the global ability in governance to effectively control new technologies would be very problematic. The general argument is therefore the following: because anthropogenic existential risk is already not sustainable today, a technological acceleration driven by the advent of quantum computers would increase existential risk.
\subsection{Concrete risk scenarios based on DQS-enabled algorithms}
While it is impossible to make accurate predictions about the long-term implications of DQS-based algorithms, we nevertheless seek to go beyond the general argument from the last paragraph. In order to do so, we will consider the two most severe (anthropogenic) existential threats to humanity: engineered pandemics and misaligned AI. These two risks are dominating the total existential risk, at least according to Ord's estimates~\cite[p.~167]{Ord2020}. Now, the risk of engineered pandemics or misaligned AI is a combination of many small factors of different causal interdependence. In general, however, it is reasonable to assume that any slight increase in the factors connected to these two anthropogenic risks, is reason for concern. 
\subsubsection{The hypothetical misuse of DQS-methods in engineered pandemics}
We have already argued that quantum computers will allow to perform certain biomedical experiments virtually by using DQS. Gain-of-function research seeks to find ways how to make an existing virus more contagious or lethal by altering the virus slightly. These experiments can produce variants of a virus that are much more dangerous than a wild type circulating in nature or among humans. Insights from gain-of-function research may help for public health authorities to adapt and prepare for these variants that can sooner arise through natural mutation. However, leaks from high-security laboratories have already been reported~\cite[]{zhou2019biosafety}, raising the question of the public health risk of such research~\cite[]{selgelid2016gain}, even if today the vast majority of researches likely weigh the pros and cons of this kind of research very carefully. \\
We could imagine that DQS makes is possible that at some point gain-of-function research will be performed partially out of the laboratory because some steps are simulated on a quantum computer. Any simplification in the required steps for gain-of-function research could then lead to more of these investigations being conducted. 
While such a scenario is, first and foremost, of hypothetical nature, it cannot be ruled out as impossible, either. With the existential risk of engineered pandemics already being considered to be among the largest risks, even a slight increase would already pose to be a significant threat.\\
In this context, we also refer to~\cite[]{jefferson2014synthetic} for an analysis of why engineering pandemics is much more difficult than sometimes depicted by non-experts including the author of this viewpoint. 
\subsubsection{DQS as a tool for misaligned artificial intelligence}
Another perspective comes from DQS being used as a tool by a misaligned artificial intelligence (AI). Misaligned AI is considered to be one of the most severe existential risk by many scholars. A well-known example of misaligned AI is the so-called paperclip AI. In this example, a powerful artificial general intelligence with superhuman capabilities might be assigned the task of maximizing the production output of paperclips. Given sufficient power, such an AI might follow precisely its given task with the unintended side effects of consuming all of the relevant raw materials on earth and also deciding to switch of all humans in order to maximize its alignment to its task~\cite[pp.~150-153]{Bostrom2014}. This example shows that a misaligned general AI might produce great harm to humans if the goals of the AI are not programmed very carefully. We argue here that DQS might be a powerful tool in the hand of a misaligned AI and thus be potentially dangerous.\\
Research and development in chemistry, pharmaceutical or material science today relies on lab work being performed manually by humans. Already without DQS, a misaligned AI could try to engage researchers for its objectives through financial incentives or social engineering. However, when significant steps of the research process can be performed virtually in much shorter time, a misaligned AI would be equipped with a much richer tool set for achieving its objective. Small-scale quantum computers are already accessible over the cloud. In a hypothetical future scenario, cloud-access to quantum computers could  allow misaligned AI with access to the internet to perform fundamental physical research. The AI could acquire new knowledge which then further advances its capabilities. This second hypothetical scenario shows that DQS as an application of quantum computing has the potential to increase existential risk by being a tool that will be used by a misaligned AI.
\subsection{Counter-argument: DQS reducing existential risk} 
As we have seen, there is a plausible trajectory by which quantum computers can amplify existential risk. However, we would like to discuss counter-arguments how quantum computing could reduce existential risk.\\
We could make the rather general argument that an arrest of technological progress is in itself an existential risk. This is because only technology will allow humanity to mitigate any existential risk that was already present before (e.g. natural existential risk)~\cite[]{bostrom2002existential}. This argument could be interpreted as a general case for the advancement of technology. However, the argument does not apply anymore when we narrow our focus to a certain field of technology. As an example we might consider knowledge about nuclear fusion. Its use for building a hydrogen bomb might have catastrophic consequences including the existential threat of a global war with these weapons of mass destruction. At the same time, there is still research being carried out to build the first large scale prototype for a civil use. Nuclear fusion might produce vasts amounts of energy, however, due to alternative green energy available, humanity does not depend on nuclear fusion to solve its energy problems. Thus, if a field of technology is expected to have worsening effect on existential risk in total, it could be reasonable to impose policy which aims at controlling technology in this field, even if this hinders the advances of useful applications of this technology.\\
For this reason, we should sharpen the counter-argument and focus on the specific potential of DQS to reduce existential risk more than it amplifies it. In the same sense as DQS might make it easier to find new dangerous chemicals, weapons or diseases, we would also expect it to speed up research in how to prevent pandemics, aiding the understanding of the cure of pathogens and helping to find new pharmaceuticals and vaccines. Indeed, DQS could potentially even lower the risks connected with gain-of-function research if it allows critical experiments to be performed only virtually instead of using actual pathogens in a laboratory where they might escape from. The conceivable scenarios how quantum computing could lower existential risk go beyond pharmaceutical applications: DQS might help us find more sustainable materials and chemical reactions. As an example, aided by DQS, humanity might discover new catalysts that will, in the long term, help to remove carbon dioxide from the atmosphere \cite[]{Reiher2017Elucidating}. This could reduce the consequences of global warming, which in a worst case scenario could be an existential risk.\footnote{Importantly, this does not imply that quantum computers will be able to help mitigate the climate catastrophe in the short or medium term. However, this is currently the time scale which is paramount in order to limit global warming to 1.5$^{\circ}$C or well below 2$^{\circ}$C~\cite[]{allan2021ipcc}.} \\ 
These points are certainly true and imply that quantum computing also holds great promises. Therefore, as with any technology, analyzing the consequences of DQS involves weighing up both risks and benefits. We would argue however, that with existential risk the technology risks should be weighted much higher than technology benefits as the very meaning of \emph{existential} implies that it does not come with a try-again opportunity. If DQS is to significantly increase existential risk in the future, finding means to control the technology becomes highly relevant. We should, however, attempt to allow the responsible use of the technology if loophole-free ways are conceivable.
\section{Conclusion} \label{sec:conclusion}
Quantum computers hold the justifiable promise of great scientific discovery and a profound impact for humanity through DQS-based algorithms. We can expect that DQS helps to significantly speed-up research timelines and lead to new scientific discoveries. However, this comes with a potential  amplification of anthropogenic existential risk if no counter-measures are undertaken. Estimating the impact of this emergent technology in the next decades to come is extremely difficult and bears very large uncertainties. Given the scope of ensuring the survival of humanity, a study of the existential risk impact of quantum computing seems of great significance.  \\
For this reason, we call for quantum risk mitigation to both assess and control the future impact of quantum computing on existential risk. We should be aware that risk mitigation suffers from the notorious Collingridge's dilemma~\cite[]{collingridge1982social}. It describes the predicament that banning technology in its early stages, before the consequences become apparent, is difficult due to limited available information. At the same time, late action risks that the technology has already dispersed and cannot be effectively controlled anymore. Therefore, regarding quantum computing, there is a strong necessity to be alert early on so that the control of the technology is still feasible. While the focus of this viewpoint is the impact of quantum computing on existential risk, subsequent investigations in this regard will likely be of benefit to both the understanding of quantum existential risk as well as technology risk of quantum computing in a broader sense. We invite for further fruitful dialogue on the technological risks of quantum computing.
\begin{acknowledgments}

\end{acknowledgments}
The author would like to thank Jordi Tura and Simon Friedrich for their valuable comments.
\bibliography{biblio}

\end{document}